\documentclass[letterpaper,12pt]{JHEP3}
\usepackage{amsmath,amssymb}
\usepackage{epsfig}
\usepackage{cite}
\usepackage[english]{babel}

\newcommand{\eq}{\begin{equation}}
\newcommand{\feq}{\end{equation}}
\newcommand{\eqn}{\begin{eqnarray}}
\newcommand{\feqn}{\end{eqnarray}}
\newcommand{\arr}{\begin{eqnarray*}}
\newcommand{\farr}{\end{eqnarray*}}

\title{Black holes and singularities in string theory}
\author{Dietmar Klemm \\
Dipartimento di Fisica dell'Universit\`a di Milano \\
Via Celoria 16, I-20133 Milano and \\
INFN, Sezione di Milano, Via Celoria 16, I-20133 Milano. \\
E-mail: \email{dietmar.klemm@mi.infn.it}}
\preprint{IFUM-807-FT \\
hep-th/0410040}

\abstract{This is a summary of a lecture I gave at the workshop
on dynamics and thermodynamics of black holes and naked singularities
at Politecnico Milano. It is directed to a non-expert audience
and reviews several ways in which string theory accounts for
black hole microstates. In particular, I give an elementary introduction
to the correspondence principle by Horowitz/Polchinski, to the
state counting for the three-charge black hole by Strominger and Vafa,
and to the recent proposal by Mathur et al.~concerning the
gravity description of black hole microstates.
The second part of the lecture is dedicated to naked singularities
and reviews an argument by Horowitz and Myers why naked
singularities are not necessarily bad. Finally, I comment on a possible
resolution of singularities in Born-Infeld type gravity theories.
}


\begin{document}

\section{Black holes}
\subsection{Black holes as thermodynamical objects}

In 1971, Hawking proved the so-called "area theorem" of black
hole physics \cite{Hawking:1971tu}. It says that under reasonable
conditions, e.~g.~no "exotic matter" with negative energy density or
the like, the total area $A$ of the event horizons of any collection of
black holes can never decrease,
\begin{equation}
\delta A \ge 0\,.
\end{equation}
This sounds curiously similar to the second law of thermodynamics,
\begin{equation}
\delta S \ge 0\,,
\end{equation}
with the area of the black hole playing the role of entropy. It might appear
that this similarity is of a very superficial nature, because the area theorem
is a mathematically rigorous consequence of general relativity, whereas the
second law of thermodynamics is believed not to be a rigorous consequence
of the laws of nature but rather a law that holds with overwhelming probability
for systems with a large number of degrees of freedom. Nevertheless, this
analogy extends to all the laws of black hole mechanics, derived by Bardeen,
Carter and Hawking in 1973 \cite{Bardeen:1973gs}, and thermodynamic principles.
This is summarized in table \ref{4laws}.

\TABLE{
    \begin{tabular}{|c||c|c|}
      \hline
      Law nr. & Thermodynamics & Black holes \\
      \hline
      \hline
      0 & $T$ constant in thermal equilibrium & Surface gravity $\kappa$ constant
      on horizon  \\
      \hline
      1 & $dE = TdS$ + work terms & $dM = \frac{\kappa}{8\pi}dA + \Omega_H dJ$ \\
      \hline
      2 & $\delta S \ge 0$ & $\delta A \ge 0$ \\
      \hline
      3 & \parbox[c]{5.5cm}{
        \centerline{Impossible to achieve $T=0$
        $\vphantom{\displaystyle1^{\displaystyle H}}$}
        \centerline{by a physical process$\vphantom{\displaystyle\frac{.}{1}}$}
      } & \parbox[c]{5.5cm}{
        \centerline{Impossible to achieve $\kappa=0$
        $\vphantom{\displaystyle1^{\displaystyle H}}$}
        \centerline{by a physical process$\vphantom{\displaystyle\frac{.}{1}}$}
      } \\
      \hline
    \end{tabular}
  \caption{Analogy between the four laws of black hole mechanics and the laws
           of thermodynamics. $M$ and $J$ denote the mass and angular momentum of
           the black hole, and $\Omega_H$ is the angular velocity of the horizon.}
  \label{4laws}
}

Physicists were thus led to ask the question if this analogy is only formal,
or if there is some deeper meaning behind. Note in this context that classically
a black hole has zero temperature, because nothing can escape from the region
behind the horizon. Hawking himself tended not to believe in a profound
meaning behind this analogy, but he had to correct his opinion when he
discovered \cite{Hawking:1974sw} that, due to quantum effects, black holes radiate
like a black body with temperature
\begin{equation}
T = \frac{\hbar \kappa}{2\pi}\,.
\end{equation}
At this point it became clear that the surface gravity $\kappa$ indeed
represents the thermodynamical temperature of the black hole, and that
the resemblance of the four laws of black hole mechanics and the
thermodynamical laws is more than just a formal analogy:
The laws of black hole mechanics are the laws of thermodynamics,
applied to black holes. In particular this implies that we should assign
the entropy
\begin{equation}
S_{BH} = \frac{A_{Hor}}{4G}\,, \label{SBH}
\end{equation}
called Bekenstein-Hawking entropy, to a black hole. This identification
raises the question what the black hole microstates are, and where
they are located. In other words, we would like to know what
the statistical mechanics of black holes is, and write (\ref{SBH})
as the logarithm of a number of microstates that are compatible with
a given macrostate, i.~e.~, with a given set ($M,J,Q$), where $Q$ stands
for the charges that can be carried by the hole.

In the remainder of this section, I will try to argue that string theory
can provide an answer to this question.

\subsection{The correspondence principle}

\begin{figure}[ht]
\begin{center}
\includegraphics[width=1.0\textwidth]{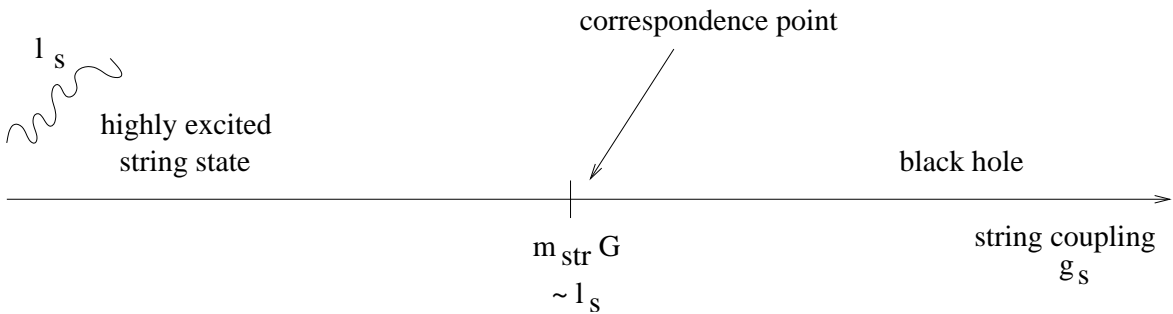}
\end{center}
\caption{\small{The Susskind-Horowitz-Polchinski correspondence
principle: A highly excited string state at low string coupling
$g_s$ turns into a black hole when one increases $g_s$, because
at a certain point the Schwarz\-schild radius
$m_{str}G \sim m_{str}g_s^2\ell_s^2$ of the string
becomes larger than the string length $\ell_s$.}
}
\label{horpol}
\end{figure}

The correspondence principle, formulated by Horowitz and Polchinski in
1996 \cite{Horowitz:1996nw}, is essentially based on Susskind's idea that
Schwarz\-schild black holes are in one-to-one correspondence with
fundamental string states \cite{Susskind:1993ws}. If one starts at week
string coupling $g_s$ with a highly excited string state, and raises
$g_s$, then also the Newton constant $G$ increases, because in four
dimensions $G$ is related to $g_s$ and the string scale $\ell_s$
by $G \sim g_s^2 \ell_s^2$. At a certain point, the Schwarz\-schild
radius of the string, $m_{str}G$, becomes larger than the string length
$\ell_s$, and the string turns into a black hole. Conversely, as one
decreases the coupling, the size of a black hole eventually becomes
less than the string scale. At this point, the metric is no longer
well-defined near the horizon, so it can no longer be interpreted as
a black hole. Susskind proposed that it should be described in terms
of some string state. The point where the black hole turns into a
string is called the correspondence point. The mechanism is represented
graphically in figure \ref{horpol}.

\vspace*{0.5cm}

\vspace*{0.5cm}

At the correspondence point one sets the string mass equal to the
black hole mass, $m_{str} = m_{bh}$. The string spectrum in flat
space is given by
\begin{equation}
m_{str}^2 \sim \frac n{\alpha'}\,, \quad n = 0,1,2,\ldots \label{strspectr}
\end{equation}
In (\ref{strspectr}), $\alpha' = \ell_s^2$ denotes the inverse of the
string tension. Combined with $m_{str}^2 \sim \ell_s^2/G^2$ this
yields
\begin{equation}
\frac{\ell_s^2}G \sim \sqrt n\,. \label{ellGn}
\end{equation}
The entropy of the four-dimensional Schwarz\-schild black hole
is given by
\begin{equation}
S_{bh} \sim \frac{r_{hor}^2}G \sim \frac{\ell_s^2}G \sim \sqrt n\,,
\end{equation}
where $r_{hor}$ is the location of the event horizon, and we used
(\ref{ellGn}) and the fact that $r_{hor} \sim \ell_s$ at
the correspondence point. On the other hand, it is well-known that
the string entropy has the same behaviour,
\begin{equation}
S_{str} \sim \sqrt n\,,
\end{equation}
so that the Bekenstein-Hawking entropy is comparable to the
string entropy. In other words, an excited string provides
the correct number of degrees of freedom to account for the entropy of
the Schwarz\-schild black hole. This approach works also in
other than four dimensions and for charged black
holes \cite{Horowitz:1996nw}.
The general idea is that, when the size of the horizon
drops below the size of a string, the black hole state
becomes a state of strings and D-branes (cf.~next subsection)
with the same charges.

Note that this calculation does not yield the correct numerical
coefficient of the Bekenstein-Hawking entropy, that's why we did
not even try to put the correct prefactors in the above equations.
It gives however the correct dependence on $n$.

\subsection{The three-charge black hole}

The first microstate
counting for black holes in string theory that reproduced also the
right numerical coefficient of the Bekenstein-Hawking entropy
was done by Strominger and Vafa \cite{Strominger:1996sh}.
They considered supersymmetric (and thus necessarily extremal and
charged) black holes. In the presence of enough supersymmetry,
there exist so-called non-renormalization theorems, which essentially
say that weak coupling results are protected from quantum corrections.
This means that the number of states one counts at weak coupling
cannot change as one increases the coupling, i.~e.~, when a black hole
forms.

\begin{figure}[ht]
\begin{center}
\includegraphics[width=0.5\textwidth]{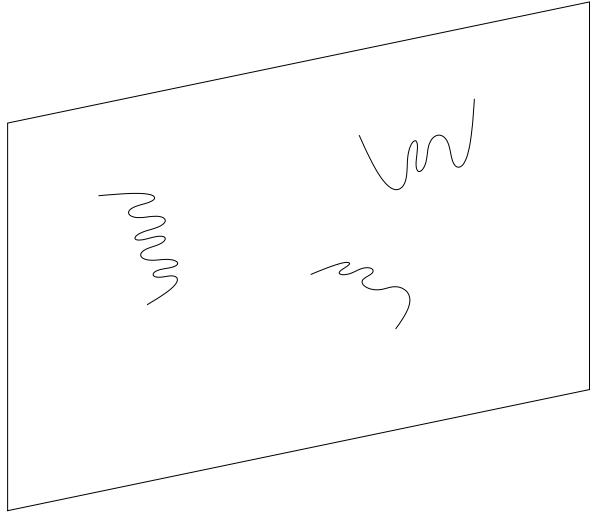}
\end{center}
\caption{\small{Dp-branes are p-dimensional hyperplanes on which open
strings can end.}
}
\label{Dbrane}
\end{figure}

In order to understand the results of \cite{Strominger:1996sh},
we need an additional input with respect to the previous subsection,
namely the concept of D-branes (where D stands for Dirichlet).
Let us therefore open a parenthesis
on D-branes. String theory is not a theory of strings alone, but it
contains also other extended objects, among these the so-called
D-branes \cite{Polchinski:1995mt}\footnote{For an introduction to D-branes
see e.~g.~\cite{Polchinski:1996na}.}, whose existence is required by string
theory dualities. Dp-branes are p-dimensional hyperplanes on which open
strings can end (cf.~figure \ref{Dbrane}).

\vspace*{0.5cm}

\vspace*{0.5cm}

The open strings that are attached to D-branes satisfy Dirichlet boundary
conditions in directions transverse to the branes (that's where the
name comes from), and Neumann boundary
conditions in directions tangential to the branes, so that they
are free to move along the branes.
The open strings ending on a Dp-brane represent the excitations of the
branes. At low energies $E \ll 1/\ell_s$, these excitations are described
by a U$(1)$ supersymmetric Yang-Mills (SYM) theory on the brane worldvolume, i.~e.~,
in $p+1$ dimensions. If we have $N$ coincident branes, the gauge group of
the SYM theory is enhanced to U$(N)$. This comes from the fact that the end
of each string has now $N$ possible D-branes on which to attach.

\begin{figure}[ht]
\begin{center}
\includegraphics[width=0.7\textwidth]{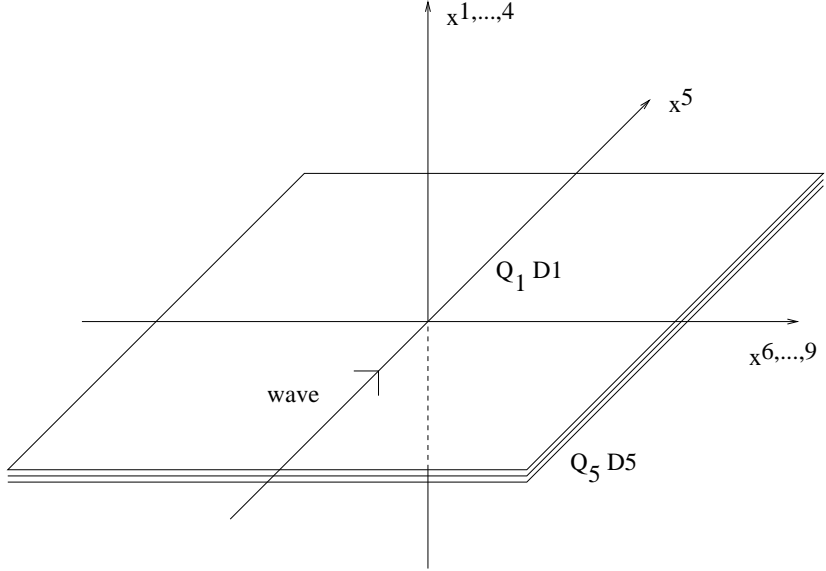}
\end{center}
\caption{\small{Configuration of $Q_5$ D5-branes along $x^5, \ldots, x^9$
and $Q_1$ D1-branes along $x^5$, with a wave propagating along the common
direction.}
}
\label{D1D5}
\end{figure}

Let us now consider the system of branes from a different point of view:
Dp-branes carry so-called Ramond-Ramond (RR) charges (they couple to a
$(p+1)$-form RR vector potential $A^{(p+1)}$, just like a particle
(which we can imagine as a "0-brane") couples to a one-form potential
$A^{(1)}$ in electromagnetism), and they have a mass. This means that
D-branes represent also sources for the gravitational field (and for
the other supergravity fields).
Now we are interested in the gravitational field of a particular configuration
of D-branes, that describes a black hole. To this end, we consider
a number $Q_5$ of coincident D5-branes along the directions $x^5, \ldots, x^9$
and $Q_1$ coincident D1-branes along $x^5$, so that their common transverse directions
are $x^1, \ldots, x^4$ (superstring theory can be formulated consistently
only in ten dimensions). Furthermore, we add a gravitational wave that moves
along the $x^5$-direction. The resulting brane plus wave configuration
is shown in figure \ref{D1D5}.

\vspace*{0.5cm}

\vspace*{0.5cm}

The gravitational field created by this source is given by
(cf.~the nice review \cite{David:2002wn} for details of the construction)
\begin{eqnarray}
ds_{10}^2 &=& f_1^{-1/2} f_5^{-1/2}[-dt^2 + (dx^5)^2 + (f_n - 1)(dt + dx^5)^2] \nonumber \\
          & & + f_1^{1/2} f_5^{1/2} dx_i\,dx^i + f_1^{1/2} f_5^{-1/2} dx_a\,dx^a\,,
\end{eqnarray}
where $i = 1,\ldots,4$, $a = 6,\ldots,9$ and
\begin{equation}
f_{1,5,n} = 1 + \frac{r^2_{1,5,n}}{r^2}\,, \qquad r^2 = (x^1)^2 + (x^2)^2 +
            (x^3)^2 + (x^4)^2\,. \label{f15n}
\end{equation}
In (\ref{f15n}), $r_{1,5}^2$ are proportional to the numbers $Q_{1,5}$ of
D1- and D5-branes, and $r_n^2$ is proportional to the momentum $N$ along the
direction $x^5$, carried by the wave (for the correct prefactors
cf.~\cite{David:2002wn}). Apart from the metric, also the
RR 3-form field strength and the dilaton of type IIB supergravity are
turned on. (Recall that the 2-form RR gauge potential couples to the
D1-branes, and its dual, which is a 6-form gauge potential, couples to the
D5-branes). This geometry preserves four of the 32 supercharges of type
IIB supergravity.

Now perform a Kaluza-Klein compactification to five dimensions along
the directions $x^5$ and $x^6, \ldots, x^9$, which we assume to
be wrapped on a circle S$^1$ and a four-torus T$^4$ respectively.
This results in the five-dimensional metric
\begin{equation}
ds_5^2 = -f^{-2/3}(r) dt^2 + f^{1/3}(r)(dr^2 + r^2 d\Omega_3^2)\,,
         \label{3charge}
\end{equation}
where $d\Omega_3^2$ denotes the round metric on the unit three-sphere and
\begin{equation}
f(r) = f_1(r) f_5(r) f_n(r)\,.
\end{equation}
The geometry (\ref{3charge}) describes the so-called three-charge black hole
(with charges $Q_{1,5}$ and $N$). In the case $r_1^2 = r_5^2 = r_n^2$, it
reduces to the extremal five-dimensional Reissner-Nordstr\"om solution.
(\ref{3charge}) has a horizon at $r=0$ and zero Hawking temperature.
The entropy of the three-charge black hole is given by
\begin{equation}
S = \frac{A_{Hor}}{4G_5} = \frac{2\pi^2 r_1 r_5 r_n}{4G_5} = 2\pi\sqrt{Q_1 Q_5 N}\,,
    \label{S3charge}
\end{equation}
where $G_5$ denotes the Newton constant in five dimensions and in the last step
we used the correct proportionality factors between $r_{1,5,n}^2$ and
$Q_{1,5},N$.

Our aim is now to reproduce the entropy (\ref{S3charge}) by counting
excitations of the D1-D5 system. In doing so, we will follow
\cite{Callan:1996dv} rather than \cite{Strominger:1996sh},
which is based on a sophisticated analysis of the cohomology of
instanton moduli spaces. Due to limitations of space, the state counting
will be presented only schematically. For the details we refer
e.~g.~to \cite{David:2002wn}.


\begin{figure}[ht]
\begin{center}
\includegraphics[width=0.7\textwidth]{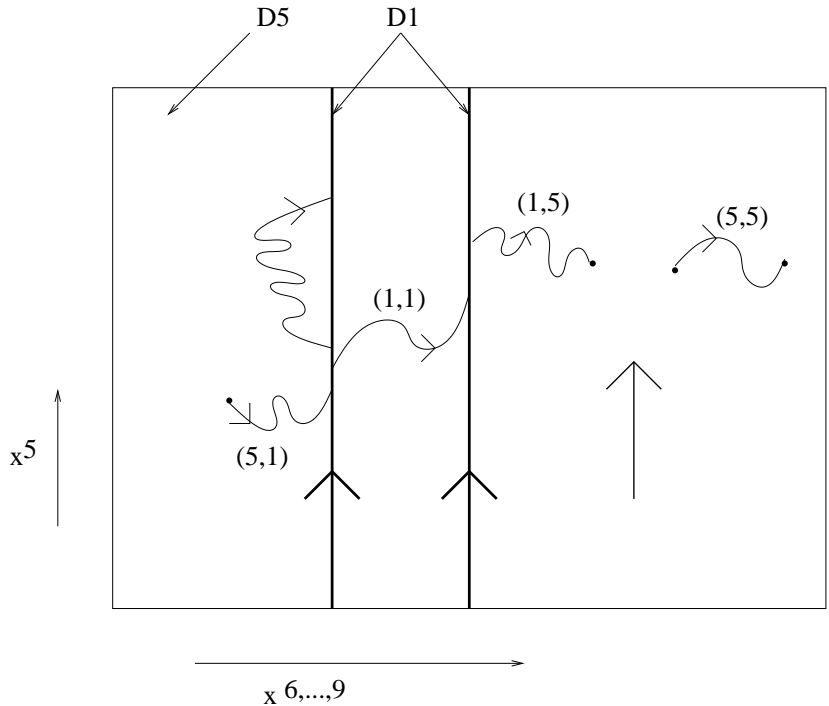}
\end{center}
\caption{\small{Various types of excitations of the D1-D5 system: (1,1)-strings that
stretch between two D1-branes, (5,5)-strings that stretch between two D5-branes, and
(1,5)- and (5,1)-strings with one end attached to a D1-brane and the other to a
D5-brane. Due to the orientation carried by the strings (denoted by an arrow in the
figure), one has to distinguish between (1,5) and (5,1). The rightmost arrow indicates
that the strings move along the $x^5$-direction, corresponding to the momentum $N$
along $x^5$.}
}
\label{excit}
\end{figure}


The various types of D1-D5 excitations are shown in figure \ref{excit}.
One can show that the low energy effective field theory describing these
excitations is a $(1+1)$-dimensional $(4,4)$
supersymmetric gauge theory with gauge group
U$(Q_1)$ $\times$ U$(Q_5)$ \cite{Maldacena:1996ky} (see also
\cite{David:2002wn} for a review). The strings of (1,5) or (5,1) type
are described by charged matter fields (corresponding to hypermultiplets)
in the fundamental representation
of U$(Q_1)$ $\times$ U$(Q_5)$. Now the presence of many open (1,5) or
(5,1) strings effectively gives an expectation value to these matter
fields, which therefore act as Higgs
fields \cite{Callan:1996dv}.
The D1-D5 system is thus described by the Higgs branch of the gauge theory.
The Higgs fields make the vector multiplets (which describe the (1,1) and the
(5,5) strings) massive, so that they can be dropped from the state counting.
One now counts the total number of bosonic degrees of freedom from the
hypermultiplets and subtracts both the number of conditions coming from the vanishing
of the superpotential (which is necessary in order to have a supersymmetric
vacuum) and the number of pure gauge degrees of freedom. This leaves $4Q_1 Q_5$
gauge invariant bosonic
degrees of freedom \cite{Maldacena:1996ky}. Supersymmetry implies then that there
are also $4Q_1 Q_5$ fermionic degrees of freedom.
Now we are interested in low energy black hole processes. For low energies
(i.~e.~, in the infrared), the above gauge theory flows to a (super-)conformal
field theory. (At very low energies, we are far below any scale in the system,
so that these scales effectively disappear and the theory becomes conformally
invariant). Two-dimensional conformal field theories are characterized by
a so-called central charge $c$ that appears in the Virasoro algebra,
and represents the number of degrees of freedom. Every (free) boson contributes
the value $1$ to the central charge, whereas a (free) fermion contributes $1/2$.
As we have $4Q_1 Q_5$ bosonic and $4Q_1 Q_5$ fermionic degrees of freedom,
the central charge reads
\begin{equation}
c = 4 Q_1 Q_5 \cdot 1 + 4 Q_1 Q_5 \cdot \frac 12 = 6 Q_1 Q_2\,.
\end{equation}
The microstates corresponding to the D1-D5 black hole are states with
eigenvalues $l_0 = N$ and $\tilde{l}_0 = 0$ of the Virasoro
generators $L_0$ and $\tilde{L}_0$ respectively.
This comes from the fact that all strings attached to the
D-branes move in the direction $x^5$ (momentum $N$), and there are no
strings moving in the opposite direction. The asymptotic number $\Omega$ of
distinct states with $l_0 = N$, $\tilde{l}_0 = 0$ is given by the Cardy
formula \cite{Cardy:1986ie} (see also \cite{Carlip:1998qw} for a nice derivation)
\begin{equation}
\Omega = \exp\left(2\pi\sqrt{\frac{cl_0}6}\right) + \exp\left(2\pi\sqrt{\frac{c\tilde{l}_0}6}
         \right) = \exp(2\pi\sqrt{Q_1 Q_5 N})\,.
\end{equation}
This yields the entropy
\begin{equation}
S = \ln\Omega = 2\pi\sqrt{Q_1 Q_5 N}\,,
\end{equation}
which is in exact agreement with the Bekenstein-Hawking entropy (\ref{S3charge}).

Callan and Maldacena showed \cite{Callan:1996dv} that, remarkably enough,
the same state counting works also for near-extremal five-dimensional black
holes. In the D-brane picture, going away from extremality means exciting also
left-moving momentum (i.~e.~, strings moving in $-x^5$-direction), and having also
antibranes. ("Anti" means that these branes carry the opposite RR charge, in the
same way in which an antiparticle carries the opposite charge of the corresponding
particle).

\begin{figure}[ht]
\begin{center}
\includegraphics[width=0.7\textwidth]{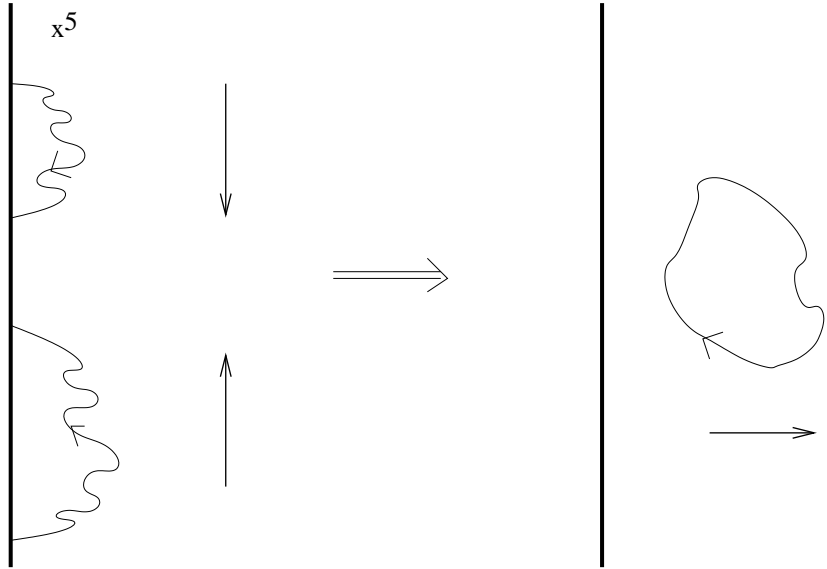}
\end{center}
\caption{\small{Hawking radiation in the D-brane/string picture: A right moving excitation
and a left-moving excitation annihilate to give a closed string that leaves the brane.}
}
\label{hawk}
\end{figure}

Let us finally see what Hawking radiation corresponds to in the D-brane/string picture.
Figure \ref{hawk} shows a typical process that leads to Hawking radiation:
A right-moving open string excitation collides with a left-moving open string
excitation to give a closed string that leaves the brane.
In \cite{Callan:1996dv}, the rate for this process was computed, and it was shown that
the radiation has a thermal spectrum at the Hawking temperature.


As we said, the geometry (\ref{3charge}) has a horizon at $r=0$, and
that's where the branes are located. If we invert the above picture
of Hawking radiation, and consider a closed string infalling towards
the horizon, i.~e.~, towards the branes, once it arrives at $r=0$,
it can split into open strings that move along the horizon.
Note that in this picture there is no information loss: Quantum states falling
into the horizon from the outside would cause a unitary evolution in the
Hilbert space of horizon states (i.~e.~, states of the D-brane system)
that "records" the infalling quantum information \cite{Callan:1996dv}.

\subsection{Gravity description of black hole microstates}

In the previous subsection we saw how to identify black hole microstates
in the D-brane picture, namely as D-brane excitations.
As one can consider the system of branes also from a different perspective,
namely as a source for the gravitational and the other supergravity fields,
we can ask the question if one can {\it see} these microstates also in
that picture (which we call the gravity side).
In other words: Can we see the black hole microstates in general relativity
(or, more generally, in supergravity)?

\begin{figure}[ht]
\begin{center}
\includegraphics[width=0.8\textwidth]{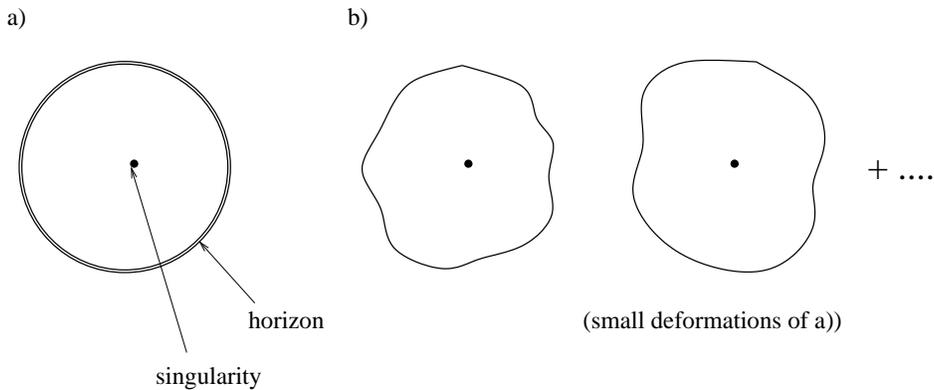}
\end{center}
\caption{\small{a) The usual picture of a black hole. b) If the microstates
represent small deformations of a) then each would itself have a horizon
and an entropy.}}
\label{hair}
\end{figure}

Early attempts to find "hair" on
black holes were based on looking for small perturbations in the metric
while demanding smoothness at the horizon. However, one found no such
perturbations. It was argued convincingly in \cite{Mathur:2003hj} that,
if we had found such hair, we would be faced with a curious difficulty:
The microstates would look like in figure \ref{hair}b), i.~e.~, like
black holes with a horizon.


\begin{figure}[ht]
\begin{center}
\includegraphics[width=0.8\textwidth]{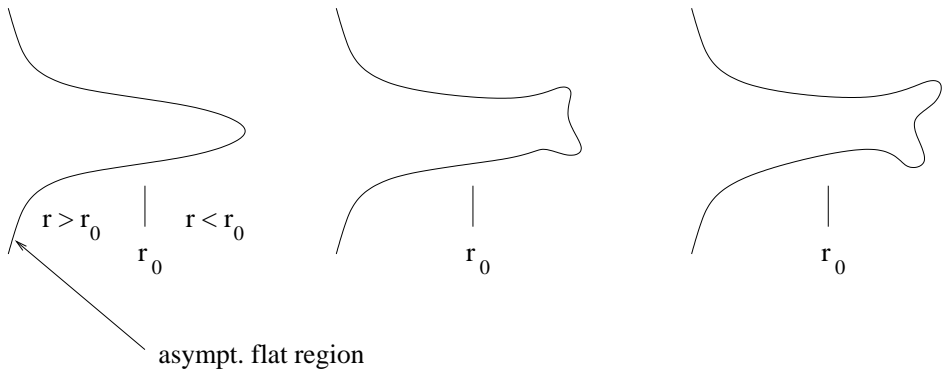}
\end{center}
\caption{\small{The metrics representing the black hole microstates
have no horizon and no curvature singularity. They all look essentially the same
for $r > r_0$, but differ from each other for $r < r_0$.}}
\label{micro}
\end{figure}

This implies that we must associate an
entropy $\approx S$ to each microstate, so we have $e^S$ configurations,
each having an entropy $\approx S$. But this makes no sense, because we
wanted the microstates to {\it explain} the entropy, not to have further
entropy themselves. Thus, if we do find the microstates in the gravity
description, then they should turn out to have {\it no horizon}
themselves. This argument led Mathur et al.~\cite{Mathur:2003hj} to the
formulation of the following requirements for black hole "hair":
\begin{enumerate}
\item There must be $e^S$ states of the hole.
\item These individual states should have no horizon and no singularity.
\item "Coarse graining" (to be explained below) over these states should give
the notion of entropy for the black hole.
\end{enumerate}
In addition, the states should carry the same charges and preserve the
same amount of supersymmetry as the hole.
In \cite{Lunin:2004uu}, various gravity microstates for the D1-D5 system were
constructed\footnote{The geometries corresponding to the ground states of the
D1-D5 system without momentum along $x^5$ (two-charge black hole)
were obtained earlier in \cite{Lunin:2001jy}.}.
The resulting metrics are rather complicated, but they share the
common feature that they have no horizon and no singularity.
Furthermore, they all look essentially the same if the radial coordinate
$r$ is larger than some value $r_0$, but differ from each other for $r < r_0$.
This is illustrated in figure \ref{micro}.

\vspace*{0.5cm}

The terminology "coarse graining" in point 3 above means now the
following: If we compute the area $A$ of the surface $r = r_0$ beyond
which the geometries look different, this should satisfy
\begin{equation}
S = \frac A{4G}\,,
\end{equation}
where $S$ is the entropy (\ref{S3charge}) of the D1-D5 black hole.
It is in this way that the notion of black hole entropy arises
in this picture.

\section{Naked singularities}
\subsection{\ldots are not necessarily bad!}

The second part of this lecture is dedicated to naked singularities.
I will first explain an argument by Horowitz and Myers \cite{Horowitz:1995ta},
that essentially states the following:
\begin{itemize}
\item Spacetime singularities play a useful role in gravitational
theories by eliminating unphysical solutions.
\item Any modification of general relativity which is completely nonsingular
cannot have a stable ground state.
\end{itemize}
To understand this, let us start from the action
\begin{equation}
S = \int d^4 x \sqrt{-g}\left[\frac R{16\pi G} + F(g_{\mu\nu},\nabla_{\mu},
    R_{\mu\nu\rho\sigma})\right]\,, \label{action}
\end{equation}
where $F$ denotes an arbitrary scalar function of the metric, the curvature
and its derivatives. (For instance in type IIB supergravity, there is a
term $F \sim {\alpha'}^3 R_{\mu\nu\rho\sigma}^4$, which comes from integrating
out massive string modes). The claim is now that the theory described by
(\ref{action}) has always solutions with unbounded curvature.
In order to see this, consider the gravitational wave
\begin{equation}
ds^2 = -du\,dv + dx_i\,dx^i + h_{ij}(u) x^i x^j\,du^2\,. \label{gravwave}
\end{equation}
This metric admits a covariantly constant null vector $l = \partial_v$. The
only nonzero component of the Ricci tensor is given by $R_{uu} = -{h^i}_i(u)$,
so (\ref{gravwave}) is a solution to general relativity if ${h^i}_i = 0$.
(Physically, the two independent components of $h_{ij}$ represent the
amplitudes corresponding to the two polarizations of the gravitational wave).
Furthermore, it is also a solution to the general theory (\ref{action}),
because the Riemann tensor of (\ref{gravwave}) is proportional to two
powers of $l_{\mu}$, and any contraction of $l_{\mu}$ vanishes, so all
second rank tensors constructed from the curvature and its derivatives
vanish as well. We can now consider the case where $h_{ij}(u)$ diverges,
which leads to a curvature singularity (in the sense of unbounded gravitational
tidal forces as the singularity is approached). This means that all extensions
of general relativity have solutions with null singularities.

Actually they must have timelike singularities as well: If, for instance, the
considered extension of general relativity regulated the negative mass
Schwarz\-schild solution, then the theory would have a regular negative
energy solution, and thus Minkowski spacetime would not be stable!
In conclusion, if we want the theory to have stable lowest energy
solutions, it must have singularities in order that one may discard what
would otherwise be pathological solutions \cite{Horowitz:1995ta}.

In general, one can interpret the appearance of a naked singularity
in a gravity solution as indicating the presence of an external source.
Hence one should not necessarily rule out such solutions as unphysical,
but rather ask if it has a reasonable physical source. Indeed, in certain
cases, one finds that the source has a reasonable physical interpretation,
in particular in string theory, where many extended sources are present.

\subsection{Born-Infeld type gravity}

It has been proposed \cite{Deser:1998rj} that curvature singularities
might be regulated in gravity theories of Born-Infeld type, in the same
way in which the divergent Coulomb potential of a point charge is regulated in
Born-Infeld theory. In order to see how this could work, let us start from
a very simple example, namely a free particle, with Newtonian Lagrangian
$\frac 12 m v^2$. Replacing this by the relativity expression
\begin{equation}
{\cal L} = mc^2 \left(1 - \sqrt{1 - \frac{v^2}{c^2}}\right)
\end{equation}
yields an upper bound on the velocity $v$, $v \le c$. This idea was used
by Born and Infeld in 1934, who felt disturbed by the infinite self-energy
of a point charge \cite{Born:1934gh}. In order to regularize this,
they introduced an upper bound on the electromagnetic field strength,
replacing ${\cal L}_0 = \frac 14 F_{\mu\nu} F^{\mu\nu}$ by
\begin{equation}
{\cal L} = b^2 \left(\sqrt{1 + \frac 1{2b^2} F_{\mu\nu} F^{\mu\nu}} - 1\right)\,,
           \label{borninf}
\end{equation}
where $b$ denotes a constant\footnote{Born and Infeld called the constant $b$, that
has dimension of a field strenth, {\it absolute field}. In the spirit of a
{\it unitarian standpoint}, in which particles of matter are considered as
singularities of fields and in which mass is a derived notion to be
expressed by field energy, they determined the value
of $b$ by equating the electromagnetic field energy of a point charge with $m_0 c^2$,
where $m_0$ denotes the electron mass. This yields $b \approx 10^{16}$
electrostatic units \cite{Born:1934gh}, which is an enormous magnitude.}.
The Lagrangian (\ref{borninf}) reduces to ${\cal L}_0$ for small fields
($|F_{\mu\nu}| \ll b$).

\begin{figure}[ht]
\begin{center}
\includegraphics[width=0.7\textwidth]{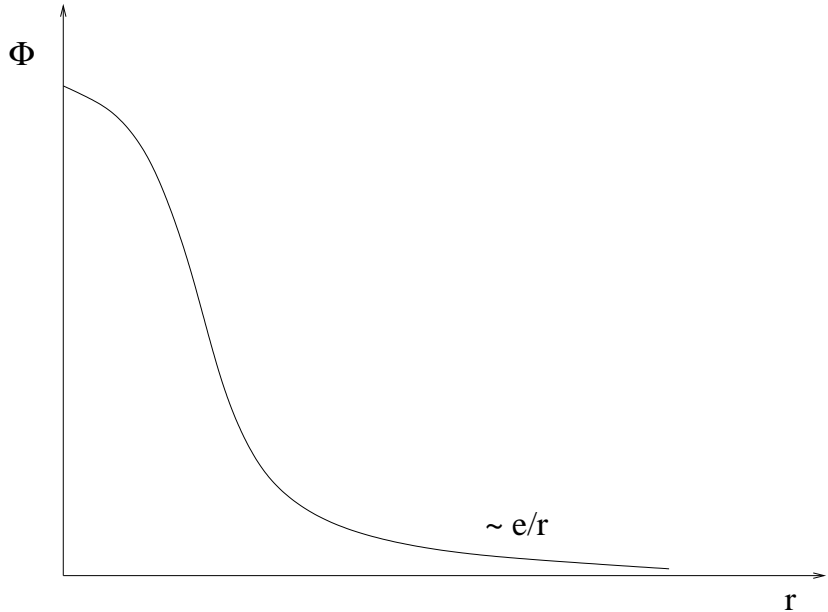}
\end{center}
\caption{\small{The potential of a point charge $e$ in Born-Infeld theory.
There is no divergence for $r \to 0$. At large distances, $\phi(r)$
goes like $e/r$.}}
\label{phi}
\end{figure}

In the new theory (\ref{borninf}), the potential of a point charge $e$ is now
given by \cite{Born:1934gh}
\begin{equation}
\phi(r) = \frac e{r_0} f\left(\frac r{r_0}\right)\,, \qquad f(x) = \int_x^{\infty}
\frac{dy}{\sqrt{1 + y^4}}\,, \qquad r_0 = \sqrt{\frac eb}\,.
\end{equation}
The behaviour of $\phi(r)$ is shown in figure \ref{phi}.
Both the potential and the electric field $E_r = -d\phi/dr$ are
finite in $r = 0$.


This example raises the question if we can do something similar
for gravity in order to eliminate curvature singularities.
A possible Born-Infeld type generalization of the Einstein-Hilbert
action would be \cite{Deser:1998rj}\footnote{Cf.~also \cite{Wohlfarth:2003ss}.}
\begin{equation}
S \sim \int d^4 x \sqrt{-\det(a g_{\mu\nu} + b R_{\mu\nu} + c X_{\mu\nu})}\,,
\end{equation}
where $a,b,c$ are constants and the tensor $X_{\mu\nu}$ is quadratic
or higher order in the curvature.
Deser and Gibbons formulated criteria that such a theory should
satisfy \cite{Deser:1998rj}:
\begin{enumerate}
\item Reduction to the Einstein-Hilbert action for small curvatures
\item Freedom of ghosts
\item Supersymmetrizability
\item Regularization of (some) singularities. (Note that, according to what was said
in the previous subsection, the negative mass Schwarz\-schild solution
should {\it not} be regularized).
\end{enumerate}
Notice that the Schwarz\-schild singularity can only be recognized from
invariants constructed with the Riemann tensor, but not from those
constructed with the Ricci tensor. This means that a gravity theory that
cures this singularity should include the full Riemann tensor in the
action.

It is at present an unsettled question if a Born-Infeld type
gravity that meets the above criteria exists.

\section{Conclusions}

In this lecture I reviewed some ways in which string theory provides
the correct black hole microstates. The general idea is that one has
a configuration of D-branes at weak string coupling that turns into
a black hole at strong coupling. The D-brane excitations are given by
open strings attached to the branes, and we saw in the particular example
of the D1-D5 system that the low energy degrees of freedom of these open
strings represent the black hole microstates and reproduce exactly the
Bekenstein-Hawking entropy. In this picture, Hawking radiation comes
from collision of left- and right-moving open string excitations on
the brane, which annihilate to give a closed string that leaves the
brane \cite{Callan:1996dv}.

I also gave a brief introduction to the recent proposal of Mathur
et al.~concerning a gravity description of black hole
microstates. We saw that these gravity microstates cannot have
an event horizon, and that coarse graining gives the notion
of black hole entropy: The geometries of the microstates differ
essentially from each other only if the radial coordinate $r$ is smaller
than some value $r_0$, where the area $A(r_0)$ of the surface $r = r_0$
satisfies $S = A(r_0)/4G$. This implies a drastic modification of our
picture of the interior of a black hole, which is not just empty space
with a singularity, but rather one has a nontrivial interior exhibiting
the degrees of freedom contributing to the entropy \cite{Mathur:2003hj}.

Finally, I commented on naked singularities and explained that
they can play a useful role by eliminating unphysical solutions,
so that one obtains a stable ground state \cite{Horowitz:1995ta}.
I tried to motivate how a gravitational theory of Born-Infeld type
might regulate {\it bad} curvature singularities, a question which
is at present unsettled.

\end{document}